\documentclass[10pt,twocolumn,english,twocolumn,aps,prd,longbibliography]{revtex4-1}
\usepackage{lmodern}

\usepackage[T1]{fontenc}
\usepackage[latin9]{inputenc}
\usepackage{babel}
\usepackage{float}
\usepackage{amsmath}
\usepackage{graphicx}
\usepackage[unicode=true,
 bookmarks=false,
 breaklinks=false,pdfborder={0 0 1},colorlinks=false,colorlinks=true,
  urlcolor=blue,
  linkcolor=blue,
  citecolor=blue]{hyperref}

\makeatletter
\usepackage{lmodern}
\usepackage[T1]{fontenc}
\usepackage{prettyref}
\usepackage{textgreek}
\usepackage{inputenc}
\usepackage{slashed}
\usepackage{hyperref}
\usepackage{bibentry}
\usepackage{color}

\makeatother

\begin{document}

\title{Complete treatment of single-photon emission in planar channeling}
\author{Tobias N. Wistisen and Antonino Di Piazza}
\affiliation{Max-Planck-Institut f{\"u}r Kernphysik, Saupfercheckweg 1, D-69117, Germany}
\begin{abstract}
Approximate solutions of the Dirac equation are found for ultrarelativistic
particles moving in a periodic potential, which depends only on one
coordinate, transverse to the largest component of the momentum of the incoming particle. As an example we employ these solutions to calculate the
radiation emission of positrons and electrons trapped in the planar
potential found between the (110) planes in Silicon. This allows us
to compare with the semi-classical method of Baier, Katkov and Strakhovenko,
which includes the effect of spin and photon recoil, but neglects
the quantization of the transverse motion. For high-energy electrons,
the high-energy part of the angularly integrated photon energy spectrum
calculated with the found wave functions differs from the corresponding
one calculated with the semi-classical method. However, for lower
particle energies it is found that the angularly integrated emission
energy spectra obtained via the semi-classical method is in fairly
good agreement with the full quantum calculation except that the positions
of the harmonic peaks in photon energy and the photon emission angles
are shifted. 
\end{abstract}
\maketitle

\section{Introduction}

Under certain circumstances, when a high-energy charged particle enters a
crystalline medium the particle dynamics is not dominated by the scattering
on single atoms, but rather by the coherent scattering on many atoms
resulting in a smooth, bound motion along crystal axes or planes \citep{Lind65}.
This motion leads to radiation emission called channeling radiation.
This has been studied both experimentally \citep{BAK1985491,BAK1988525,PhysRevLett.43.1723,1402-4896-24-3-015,ANDERSEN1982209,PhysRevB.31.68,PhysRevLett.42.1148,PhysRevD.86.072001,RevModPhys.77.1131,Wistisen2018experimental,PhysRevLett.112.254801,Wistisen2017}
and theoretically \citep{KUMAKHOV197617,kumakhov1977theory,1402-4896-24-3-015,SAENZ198190}.
Channeling radiation from high-energy electrons/positrons represents
one of the few experimental realizations of non-perturbative and non-linear
problems in quantum electrodynamics, where the field strength experienced
by the particle in its rest frame approaches the Schwinger field $E_{\text{cr}}=1.3\times10^{16}$
V/cm. The only other experiments where this has been realized were
the SLAC laser experiment \citep{bula1996observation} and the recently
reported experiments \citep{PhysRevX.8.011020,PhysRevX.8.031004},
although with smaller quantum non-linearity parameter. The quantum
nonlinearity parameter is the ratio of the field strength experienced
by a charged particle in its rest frame and the Schwinger field strength.
While the quantum treatment of radiation emission in a laser field
can be fully treated using the Volkov state \citep{Ritus,PhysRevA.80.053403,dinu2018single,PhysRevA.83.022101,PhysRevA.83.032106,PhysRevLett.110.070402,PhysRevA.91.033415},
no such full treatment has been presented for channeling radiation.
The theory of channeling radiation so far has consisted of a quantum
approach involving wave functions at low particle energies, but which
neglects spin, photon recoil, and non-dipole transitions in the emission
process, because these effects are not important at low energies as
compared to the effect of the quantization of the energy levels in
the potential. Such an approach can be found, for example, in Refs.
\citep{ANDERSEN1982209,LERVIG1967481,AndersenBloch}. For higher particle
energies, the theory relies on the semi-classical operator method
by Baier et al. \citep{baier1968processes}, which then includes the
effects of spin and photon recoil, but neglects the quantization of
the transverse motion. The semi-classical method is relatively easy
to implement numerically and this explains why this method is often
employed for numerical calculations in crystal channeling, and crystalline
undulators \citep{korol2013channeling,Bandiera201544,PhysRevA.86.042903}.
\begin{figure}[t]
\includegraphics[width=1\columnwidth]{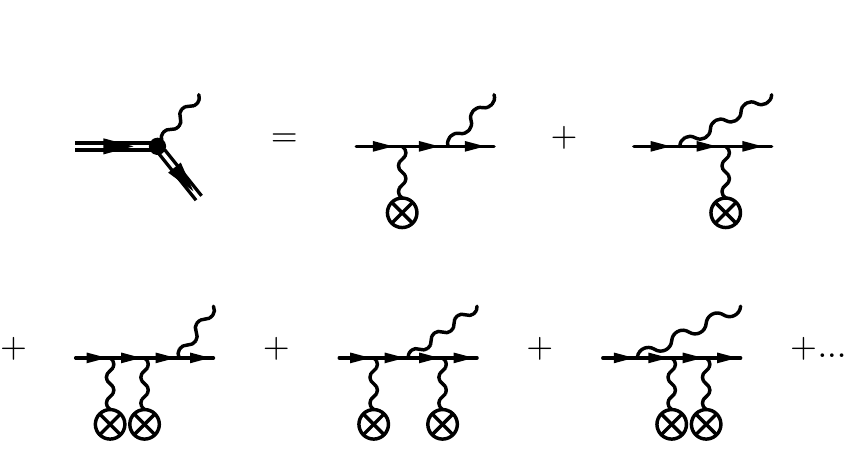} \caption{The Feynman diagrams corresponding to the process under study. The
double fermion lines correspond to the electron solutions of the Dirac
equation in the background field of the inter planar crystal potential,
which in the perturbative picture corresponds to including all orders
of interactions with this background field.}
\end{figure}
In many of the original theoretical works on channeling radiation
the connection between the Dirac equation and a Schr{\"o}dinger like equation
was seen but then various approximations were employed, such as a
simplified potential yielding analytical solutions, the dipole approximation
in the calculation of the radiation and/or the assumption that $\omega/\varepsilon$
is small, where $\omega$ is the energy of the emitted photon and $\varepsilon$ the initial energy of the radiating particle, see \citep{Kumakhov1977,zhevago1978emission,beloshitskii1978quantum,bazylev1980spectral,Kimball198569}.
See also \citep{SHULGA201716,KOZLOV20104690,SHULGA20161,DIPIAZZA20171,ABDRASHITOV20183141,PhysRevLett.115.025504,PhysRevLett.119.024801,KOROL201641,PhysRevLett.111.255502,BACKE201337,PhysRevLett.114.074801,PhysRevAccelBeams.19.071001,PhysRevLett.110.115503} for recent advancements on the subject of crystal channeling. In this paper, we present the calculation for planar channeling,
based on wave functions that are approximate solutions of the Dirac
equation in the realistic Doyle-Turner model of the periodic crystal
potential and calculate the single-photon radiation emission without
any of the mentioned approximations. Therefore, we now include all
relevant quantum effects, which in some cases yield differences as
compared to the semi-classical theory. In particular, the fact that
the radiation emission stems from transitions between discrete bound
states between the planes, is properly taken into account. We investigate
an example of $20$ GeV planar channeled electrons as in this case
one can have significant radiation emission from bound states with
low quantum number. In this case the quantization of the motion is
important and at the same time photon energies comparable to the electron
energy are emitted, such that spin and photon recoil effects are also
important. In addition, for very high particle energies, transitions
from bound states with high to low quantum numbers become more likely.
Therefore, the high-energy part of the spectrum is different from
that obtained via the semi-classical model, as states with low quantum
numbers are only approximately accounted for by the latter model.
As an example, we will see this effect for planar channeled electrons
with initial energy of $250$ GeV and $1$ TeV.

We use units where $\hbar=c=1$, $\alpha=e^{2}$, with $e$ being
the positron charge, and the Feynman slash notation such that $\slashed a=a_{\mu}\gamma^{\mu}$,
where $\gamma^{\mu}$ are the Dirac gamma matrices and $a^{\mu}$
an arbitrary four-vector. We adopt the metric tensor $\eta^{\mu\nu}=\text{diag}(+1,-1,-1,-1)$.
Below, when the term `particle' is employed, it will refer to either
an electron or to a positron.

\section{Wave functions\label{sec:Wave-functions}}

In Ref. \citep{PhysRevA.98.022131} we discussed the Dirac equation
in a parabolic potential in the regime where $\xi\ll\gamma$, where
$\xi$ is the typical transverse momentum (here $p_{y}$) divided
by the electron mass $m$ and $\gamma$ is the Lorentz gamma factor
of the incoming particle. For channeling we have that $\xi=\gamma\theta_{c}$
where $\theta_{c}=\sqrt{2V_{0}/\varepsilon}$ is the critical angle
for channeling \citep{Lind65}, and so the requirement of validity
reduces to $\theta_{c}\ll1$ or $\varepsilon\gg V_{0}$. For Silicon
$V_{0}=22.7$ eV and we are interested in ultra-relativistic particles
with energy on the scale of GeV, such that this approximation is safely
applicable. Note that at $\xi\gg1$ the overall emission angle ($\sim 2\xi/\gamma$) is much larger than the instantaneous one ($\sim 1/\gamma$) and this corresponds to the regime where the local constant
field approximation becomes applicable \citep{Baier1998}. When $\xi$
is small, however, the dipole approximation may be used \citep{Baier1998}.
Another important parameter is the so-called quantum nonlinearity
parameter already mentioned in the introduction and defined by 
\begin{equation}
\chi=\frac{e\sqrt{|(F^{\mu\nu}p_{\nu})^{2}|}}{m^{3}}.\label{eq:chiparam}
\end{equation}
When $\chi$ becomes on the order of unity, the effects of particle spin and
photon recoil become important in the radiation emission process.
The quantum description of radiation emission using wave functions
as seen in Refs. \citep{ANDERSEN1982209,LERVIG1967481,AndersenBloch}
is valid when $\chi\ll1$, because spin effects and recoil are neglected,
and when $\xi\ll1$, because the magnetic field in the particle's rest
frame is neglected and only the dipole matrix element is calculated.
In the present paper we treat the problem from the laboratory frame,
and the field can therefore be described solely by an electrostatic
potential $\varphi(y)$. We then assume that the largest component
of the particle momentum is along the $x$ direction. In Ref. \citep{PhysRevA.98.022131} (and see Appendix A)
we found that the positive-energy solutions of the Dirac equation,
to leading order in $\xi/\gamma$, can be written as (we set the quantization
volume $V=1$) 
\begin{equation}
\psi(x)=\frac{1}{\sqrt{2\varepsilon}}e^{i(p_{x}x+p_{z}z-\varepsilon t)}U(y),\label{eq:electronwavefunc}
\end{equation}
and $U(y)$ is given by 
\begin{equation}
U(y)=\sqrt{\varepsilon+m}\left(\begin{array}{c}
\boldsymbol{s}\\
\frac{\boldsymbol{\sigma}\cdot\tilde{\boldsymbol{p}}}{\varepsilon+m}\boldsymbol{s}
\end{array}\right)I(y),\label{eq:Uspinor}
\end{equation}
where $\tilde{\boldsymbol{p}}=\left(p_{x}-e\varphi(y),-i\frac{d}{dy},p_{z}\right)$,
$\boldsymbol{s}$ is a two component vector describing the spin, which
we can choose as either $\left(\begin{array}{cc}
1 & 0\end{array}\right)^{T}$ or $\left(\begin{array}{cc}
0 & 1\end{array}\right)^{T}$, corresponding to spin-up and spin-down respectively. 
Note that in Eq. (\ref{eq:Uspinor}) we approximated 
\begin{equation}
\frac{\boldsymbol{\sigma}\cdot\boldsymbol{p}}{\varepsilon+m+e\varphi(y)}\simeq\frac{\boldsymbol{\sigma}\cdot\tilde{\boldsymbol{p}}}{\varepsilon+m},
\end{equation}
where $\boldsymbol{p}=\left(p_{x},-i\frac{d}{dy},p_{z}\right)$, see also Appendix A.
As already
mentioned, we have taken $p_{x}$ to be the longitudinal direction,
that is, in the initial state $p_{z}=0$, $p_{x}>0$ and $p_x\simeq\varepsilon\gg m\xi$, where $\varepsilon=m\gamma$ is the initial (constant) energy of the particle. The
function $I(y)$ is the solution of the equation 
\begin{equation}
\left[-\frac{1}{2\varepsilon}\frac{d^{2}}{dy^{2}}-e\varphi(y)\right]I(y)=\frac{\varepsilon^{2}-p_{x}^{2}-p_{z}^{2}-m^{2}}{2\varepsilon}I(y).\label{eq:schrodeq}
\end{equation}
For $\varphi(y)$ we will use the Doyle-Turner model \citep{Baier1998,Doyle,avakian1982,Moller1995403},
chosen as symmetric around $y=0$. Clearly Eq. (\ref{eq:schrodeq})
corresponds to an eigenvalue problem in the form we are accustomed
to from atomic physics. In a crystal the potential $\varphi(y)$ is
periodic with the period of the inter planar distance which we will
denote as $d_{p}$. Now, because of this periodicity the solution
can be written as a Bloch wave such that 
\begin{equation}
I(y)=e^{ik_{B}y}u_{k_{B}}(y),\label{eq:Idef}
\end{equation}
and where $u_{k_{B}}(y)$ is also periodic with period $d_{p}$. The
quantity $k_{B}$ is the Bloch momentum, which can be taken to be
in the interval $0\leq k_{B}<k_{0}$, $k_{0}=\frac{2\pi}{d_{p}}$.
It then follows from Bloch's theorem that these solutions form an
orthogonal and complete set of solutions of Eq. (\ref{eq:schrodeq})
\citep{ashcroft1976solid} (see Appendix B for
a proof that the resulting solutions of the Dirac equation are also
orthonormal within our level of approximation). Now, we are interested
in the solution $u_{k_{B}}(y)$ which is the nontrivial part of the
wave function. Inserting $I(y)$ of Eq. (\ref{eq:Idef}) into Eq.
(\ref{eq:schrodeq}) gives us the equation governing $u_{k_{B}}(y)$
\begin{flalign}
\left[-\frac{1}{2\varepsilon}\left(\frac{d^{2}}{dy^{2}}+2ik_{B}\frac{d}{dy}-k_{B}^{2}\right)+q\varphi(y)\right]u_{k_{B}}(y)\nonumber \\
=\frac{\varepsilon^{2}-p_{x}^{2}-p_{z}^{2}-m^{2}}{2\varepsilon}u_{k_{B}}(y).\label{eq:ukbdiffeq}
\end{flalign}
The periodicity of $u_{k_{B}}(y)$ implies that it can be written
as the Fourier series 
\begin{equation}
u_{k_{B}}(y)=\sum_{j}c_{j}e^{ijk_{0}y}.\label{eq:planewaveu}
\end{equation}
To write any periodic function as a Fourier series, the sum includes
infinitely many terms. However, for the numerical implementation we
are restricted to reducing the series to a finite sum. When the number
of basis vectors is increased, it is found that the lowest lying states
do converge, and therefore one only needs enough basis elements, that
the sum describing the states of interest has converged to a fixed
degree of accuracy. To ensure normalization we should have $\sum_{j}\left|c_{j}\right|^{2}=1$
(see Appendix B). It is now clear that this is an eigenvalue
problem where the quantized eigenvalue is 
\begin{equation}
E_{n}=\frac{\varepsilon^{2}-p_{x}^{2}-p_{z}^{2}-m^{2}}{2\varepsilon},\label{eq:energyquant}
\end{equation}
where $n$ is the quantum number corresponding to the value of this
energy in ascending order and where $n=0$ corresponds the ground
state. This equation leads to a quantization of, e.g., $p_{x}$. That
is, if $\varepsilon$ is fixed by the incoming energy of the particle,
a larger quantum number $n$ corresponds to a smaller value of $p_{x}$
in order to accommodate for the larger transverse momentum in the
$y$ direction. From this relation it is also clear that the quantity
$2\varepsilon E_{n}$ is related to the square of the momentum in
the $y$ direction, $p_{y}^{2}$. The coefficients $c_{j}$ are found
by solving the matrix eigenvalue problem obtained by inserting Eq.
(\ref{eq:planewaveu}) in Eq. (\ref{eq:ukbdiffeq}), by multiplying
by $e^{-ilk_{0}y}/d_{p}$, and by integrating over $y$ from $-d_{p}/2$
to $d_{p}/2$ to exploit orthogonality: 
\begin{align}
 & \sum_{j}\frac{1}{2\varepsilon}\left[jk_{0}+k_{B}\right]^{2}\delta_{j,l}c_{j}\nonumber \\
 & +\sum_{j}c_{j}\frac{1}{d_{p}}\int q\varphi(y)e^{i(j-l)k_{0}y}dy\nonumber \\
 & =\sum_{j}\frac{\varepsilon^{2}-p_{x}^{2}-p_{z}^{2}-m^{2}}{2\varepsilon}\delta_{j,l}c_{j},\label{eq:Matrixeq}
\end{align}
where $q=e$ for a positron and $q=-e$ for an electron. With these
results taken into consideration, we now see that we can write the
function $U(y)$ in terms of the coefficients $c_{j}$ such that 
\begin{equation}
U(y)=\sum_{j}c_{j}\boldsymbol{S}_{j}e^{i(jk_{0}+k_{B})y},\label{eq:Uplanewave}
\end{equation}
where 
\begin{equation}
\boldsymbol{S}_{j}=\sqrt{\varepsilon+m}\left(\begin{array}{c}
\boldsymbol{s}\\
\frac{\boldsymbol{\sigma}\cdot\boldsymbol{p}_{j}}{\varepsilon+m}\boldsymbol{s}
\end{array}\right),\label{eq:bigS}
\end{equation}
and where 
\begin{equation}
\boldsymbol{p}_{j}=\left(p_{x}+E_{n}-\frac{(jk_{0}+k_{B})^{2}}{2\varepsilon},jk_{0}+k_{B},p_{z}\right).
\end{equation}
In order to write the momentum in this form we have replaced the term
with the potential by exploiting Eq. (\ref{eq:ukbdiffeq}). Before
calculating the radiation emission probability we will calculate the
expectation value of the momentum in the $y$ direction, as this will
provide insight on how the momentum relates to the quantum number
$k_{B}$. By inserting our wave function from Eq. (\ref{eq:electronwavefunc})
in 
\begin{equation}
\left\langle p_{y}\right\rangle =\int_{-\infty}^{\infty}\psi^{\dagger}(x)\left(-i\frac{d}{dy}\right)\psi(x)d^{3}x,\label{eq:momentumdef}
\end{equation}
we find that this becomes simply 
\begin{equation}
\left\langle p_{y}\right\rangle =\sum_{j}\left|c_{j}\right|^{2}(jk_{0}+k_{B}),\label{eq:momentumexpect}
\end{equation}
to leading order in our approximation, see Appendix C.

\section{Single photon emission\label{sec:Single-photon-emission}}

\begin{figure}[t]
\includegraphics[width=1\columnwidth]{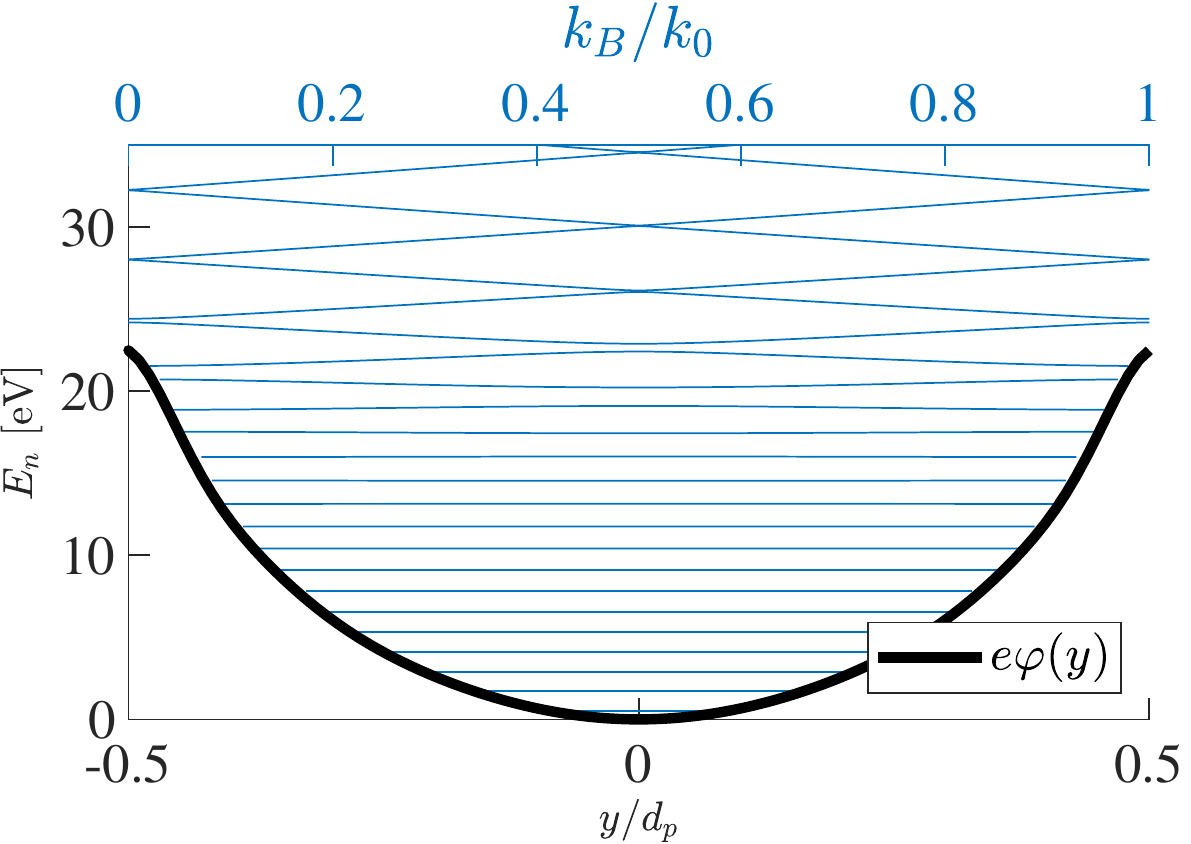} \caption{The potential $\varphi(y)$ along with the positron energy bands in
the first Brillouin zone for the energy $\varepsilon=100$ MeV. This
energy was chosen to have a reasonable number of levels so that the
plot is not cluttered.\label{fig:figpot}}
\end{figure}

We will now derive the single photon emission probability. The leading-order
$S$-matrix element for the emission of a single photon by an electron
moving inside the potential $\varphi(y)$ is given by 
\begin{flalign}
S_{i\rightarrow f}^{(1)} & =-i\sqrt{\frac{4\pi}{2\omega}}\int_{-\infty}^{\infty}d^{4}x\bar{\psi}_{f}(x)e\slashed{\epsilon}^{*}e^{ikx}\psi_{i}(x),
\end{flalign}
and the emission probability is then 
\begin{equation}
dP_{i\rightarrow f}^{(1)}=\left|S_{i\rightarrow f}^{(1)}\right|^{2}\frac{d^{3}p_{f}}{(2\pi)^{3}}\frac{d^{3}k}{(2\pi)^{3}}.\label{eq:propability}
\end{equation}
By inserting our wave functions and integrating over the coordinates,
which provide energy-momentum conservation delta functions, we have
that 
\begin{flalign}
S_{i\rightarrow f}^{(1)}= & -i\mathcal{M}_{i\rightarrow f}(2\pi)^{4}\delta(\varepsilon_{f}+\omega-\varepsilon_{i})\nonumber \\
 & \times\delta(p_{x,i}-k_{x}-p_{x,f})\delta(p_{z,i}-k_{z}-p_{z,f})\nonumber \\
 & \times\delta(k_{B,i}-k_{y}-k_{B,f}-n_{B}k_{0}),
\end{flalign}
where $n_{B}$ is the integer such that $0\leq k_{B,f}<k_{0}$, henceforth
denoted as the first Brillouin zone (FBZ) (see Appendix D
for details on this derivation). In the above expression of the $S$-matrix
element, we defined the reduced matrix element 
\begin{equation}
\mathcal{M}_{i\rightarrow f}=e\sqrt{\frac{4\pi}{2\omega}}\frac{1}{2\sqrt{\varepsilon_{f}\varepsilon_{i}}}\sum_{j}c_{n_{B}+j,f}^{*}c_{j,i}\bar{\boldsymbol{S}}_{n_{B}+j,f}\slashed\epsilon_{k}\boldsymbol{S}_{j,i},\label{eq:Mtilde}
\end{equation}
where $\boldsymbol{S}_{j,i}$ corresponds to the initial state, $c_{j,i}$
is the coefficient with index $j$ corresponding to the initial state
$i$. It is therefore seen that one must solve the matrix problem
of Eq. (\ref{eq:Matrixeq}) for $\varepsilon_{f}=\varepsilon_{i}-\omega$
several times, as the final state depends on the energy of the emitted
photon $\omega$ and the Bloch momentum $k_{B,f}$ of the final electron
(see Appendix D for additional details especially
on why $\mathcal{M}_{i\rightarrow f}$ reduces to a single sum over
$j$). We find that the rate corresponding to Eq. (\ref{eq:propability}),
when dividing by total interaction time $T_{\text{int}}$ arising
from squaring the energy delta function, is given by

\begin{flalign}
dW_{i\rightarrow f}^{(1)} & =\frac{1}{(2\pi)^{2}}\left|\mathcal{M}_{i\rightarrow f}\right|^{2}\delta(\varepsilon_{f}+\omega-\varepsilon_{i})d^{3}k.\label{eq:Onephotemission}
\end{flalign}
As is usually the case when dealing with calculations concerning ultra-relativistic
particles it is pertinent to consider cancellations between the large
terms in the expression in the delta function, $\varepsilon_{f}+\omega-\varepsilon_{i}$,
because the relevant transverse energies $E_{n}$, comparable to the
potential depth, are much smaller than the whole energy of the particle
(recall that the former are of the order of several eV, whereas the
latter is of the order of GeV). For this it is useful to consider
the quantity $\Delta$ which we define via the equation $\varepsilon=p_{x}+\Delta$.
Inserting this into the equation for our eigenvalues, Eq. (\ref{eq:energyquant}),
we obtain that to leading order in $E_{n}/\varepsilon$, 
\begin{equation}
\Delta=E_{n}+\frac{p_{z}^{2}+m^{2}}{2\varepsilon}.\label{eq:delta-1}
\end{equation}
Defining $\boldsymbol{k}=\omega(\text{cos}\theta,\text{sin}\theta\text{cos}\varphi,\text{sin}\theta\text{sin}\varphi)$
we can rewrite the argument of the remaining energy delta function
as (note that in the present problem it is convenient to use the $x$
axis as polar axis) 
\begin{flalign}
f(\theta) & =\varepsilon_{f}+\omega-\varepsilon_{i}\nonumber \\
 & =E_{n_{f}}-E_{n_{i}}+\frac{p_{z,f}^{2}+m^{2}}{2\varepsilon_{f}}-\frac{p_{z,i}^{2}+m^{2}}{2\varepsilon_{i}}\nonumber \\
 & +p_{x,f}-p_{x,i}+\omega\nonumber \\
 & \simeq E_{n_{f}}-E_{n_{i}}+\frac{m^{2}}{2\varepsilon_{f}}-\frac{m^{2}}{2\varepsilon_{i}}\nonumber \\
 & +\frac{\omega\theta^{2}}{2}\left(1+\frac{\omega\text{sin}^{2}\varphi}{\varepsilon_{f}}\right).\label{eq:ftheta}
\end{flalign}
%Now we may use that $\delta(\varepsilon_{f}+\omega-\varepsilon_{i})=\frac{1}{|f'(\theta_{0})|}\delta(\theta-\theta_{0})$
where $\theta_{0}$ is the positive solution to $f(\theta)=0$. When
considering bound states, the energies $E_{n}$ become nearly independent
of $k_{B}$, as seen in Fig. (\ref{fig:figpot}). In this case one
can isolate the emission angle as 
\begin{equation}
\theta_{0}=\sqrt{\frac{2\left(E_{n_{i}}-E_{n_{f}}\right)+\frac{m^{2}}{\varepsilon_{i}}-\frac{m^{2}}{\varepsilon_{f}}}{\omega\left(1+\frac{\omega\text{sin}^{2}\varphi}{\varepsilon_{f}}\right)}},\label{eq:Quantumangle}
\end{equation}
and the threshold for a given transition is given by the condition
where the numerator vanishes, which gives us that 
\begin{equation}
\omega_{\text{th,q}}=\varepsilon_{i}\left(1-\frac{1}{1+\frac{2\varepsilon_{i}\left(E_{n_{i}}-E_{n_{f}}\right)}{m^{2}}}\right).\label{eq:quantumthreshold}
\end{equation}

\section{Semi-classical method\label{sec:Semi-classical-method}}

In order to compare with the classical theory we will derive a formula
for the emitted power in the case of a periodic transverse motion,
and then in the end this can be turned into a semi-classical formula
by comparing with the expression of Baier et al. {[}see Eq. (\ref{eq:Baier formula})
below{]}. Classically the emitted energy per unit frequency and solid
angle can be written as \citep{Jack99,PhysRevD.90.125008} 
\begin{equation}
\frac{d^{2}I^{\text{Cl}}}{d\omega d\Omega}=\frac{e^{2}\omega^{2}}{4\pi^{2}}\left|\int_{-\infty}^{\infty}\left(\boldsymbol{n}-\boldsymbol{v}\right)e^{i(kx)}dt\right|^{2},\label{eq:classicalformula}
\end{equation}
where $k^{\mu}=(\omega,\boldsymbol{k})$ is the four-momentum of the
emitted radiation, where $x^{\mu}=(t,\boldsymbol{r})$ is the four-position
of the particle, $\boldsymbol{v}$ its velocity, and $\boldsymbol{n}=\boldsymbol{k}/\omega$.
In the semi-classical formalism the same quantity is given by \citep{Baier1998,baier1968processes,PhysRevD.90.125008}
\begin{align}
\frac{d^{2}I}{d\omega d\Omega} & =\frac{e^{2}\omega'^{2}}{4\pi^{2}}\left(\frac{\varepsilon^{2}+\varepsilon'^{2}}{2\varepsilon^{2}}\left|\int_{-\infty}^{\infty}\left(\boldsymbol{n}-\boldsymbol{v}\right)e^{i(k'x)}dt\right|^{2}\right.,\nonumber \\
 & \left.+\frac{\omega^{2}m^{2}}{2\varepsilon^{4}}\left|\int_{-\infty}^{\infty}e^{i(k'x)}dt\right|^{2}\right),\label{eq:Baier formula}
\end{align}
where $\varepsilon'=\varepsilon-\omega$, $\omega'=\omega\varepsilon/\varepsilon'$,
and $k'=k\varepsilon/\varepsilon'$ (this result holds in the case
when the sum over final particle spins and photon polarizations and
the average over initial spins are taken). As it was shown in Refs.
\citep{PhysRevD.90.125008,PhysRevD.92.045045} we need only to calculate
the transverse components of the integrand for ultra-relativistic
particles as the longitudinal component is suppressed by at least
a factor of $1/\gamma$ in comparison. Therefore, for the classical
formula we have to calculate 
\begin{flalign}
\int_{-\infty}^{\infty}e^{i\omega(t-\boldsymbol{n}\cdot\boldsymbol{r})}dt= & \int_{-\infty}^{\infty}e^{i\omega(t-\boldsymbol{n}_{\bot}\cdot\boldsymbol{r}_{\bot}-n_{x}x(t))}dt.
\end{flalign}
Since most of the radiation is emitted in the forward direction for
ultra-relativistic particles, we may perform the small-angle expansion
and so we write $n_{x}=1-\frac{\theta^{2}}{2}$ and $x(t)=\left\langle v_{x}\right\rangle t+\delta x(t)$,
where we exploit the fact that the motion is quasi-periodic. For this
reason, the quantity $\delta x(t)$ is a periodic function. By using
the fact that $v^{2}$ is approximately conserved we have that $\left\langle v_{x}\right\rangle \simeq1-\frac{1}{2\gamma^{2}}-\frac{\boldsymbol{v}_{\bot}^{2}}{2}$,
see also Ref. \citep{Jack99}. Inserting this result in the above
expression and canceling the large terms we obtain

\begin{align}
 & \int_{-\infty}^{\infty}e^{i\omega(t-\boldsymbol{n}\cdot\boldsymbol{r})}dt\nonumber \\
= & \int_{-\infty}^{\infty}e^{i\omega\Big[t\Big(\frac{\theta^{2}}{2}+\frac{\left\langle \boldsymbol{v}_{\perp}^{2}\right\rangle }{2}+\frac{1}{2\gamma^{2}}\Big)-\boldsymbol{n}_{\perp}\cdot\boldsymbol{r}_{\perp}-\delta x\Big]}dt.
\end{align}
Now, if $\left\langle \boldsymbol{v}_{\perp}\right\rangle $ is zero,
i.e., we have chosen the coordinate system where this is the case,
we can exploit that $\boldsymbol{r}_{\perp}$ and $\delta x$ are
periodic and write $\int_{-\infty}^{\infty}f(t)dt=\sum_{n}\int_{nT}^{(n+1)T}f(t)dt$,
where $T$ is the period of the motion, and change variable at which
point the Dirichlet kernel appears, which can be replaced by a sum
of delta functions. Thus, we obtain 
\begin{align}
 & \int_{-\infty}^{\infty}e^{i\omega(t-\boldsymbol{n}\cdot\boldsymbol{r})}dt\nonumber \\
= & C\left(\omega,\theta\right)\sum_{n}\delta\left(\frac{\omega}{2\gamma^{2}\omega_{0}}\left(1+\gamma^{2}\theta^{2}+\gamma^{2}\left\langle \boldsymbol{v}_{\perp}^{2}\right\rangle \right)-n\right),
\end{align}
where $\omega_{0}=\frac{2\pi}{T}$ and where we introduced the function
\begin{equation}
C\left(\omega,\theta\right)=\int_{0}^{T}e^{i\omega\Big[t\Big(\frac{\theta^{2}}{2}+\frac{\left\langle \boldsymbol{v}_{\perp}^{2}\right\rangle }{2}+\frac{1}{2\gamma^{2}}\Big)-\boldsymbol{n}_{\perp}\cdot\boldsymbol{r}_{\perp}-\delta x\Big]}dt.\label{eq:C}
\end{equation}
Analogously, we introduce the quantity 
\begin{equation}
\boldsymbol{D}_{\bot}\left(\omega,\theta\right)=\int_{0}^{T}\boldsymbol{v}_{\perp}e^{i\omega\Big[t\Big(\frac{\theta^{2}}{2}+\frac{\left\langle \boldsymbol{v}_{\perp}^{2}\right\rangle }{2}+\frac{1}{2\gamma^{2}}\Big)-\boldsymbol{n}_{\perp}\cdot\boldsymbol{r}_{\perp}-\delta x\Big]}dt.\label{eq:D}
\end{equation}
For the integral with the velocity, completely analogous steps can
be taken and therefore we have that 
\begin{align}
 & \int_{-\infty}^{\infty}\boldsymbol{v}_{\perp}e^{i\omega(t-\boldsymbol{n}\cdot\boldsymbol{r})}dt\nonumber \\
= & \boldsymbol{D}_{\bot}\left(\omega,\theta\right)\sum_{n}\delta\left(\frac{\omega}{2\gamma^{2}\omega_{0}}\left(1+\gamma^{2}\theta^{2}+\gamma^{2}\left\langle \boldsymbol{v}_{\perp}^{2}\right\rangle \right)-n\right).
\end{align}
By inserting these quantities into the classical formula and using
that the delta function squared gives us the delta function again
with a factor of $\Delta\varphi/2\pi=T_{\text{int}}\omega_{0}/2\pi$.
Therefore, we obtain that the emitted energy per unit time $T_{\text{int}}$
is given by 
\begin{flalign}
\omega dW^{\text{Cl}} & =\frac{e^{2}\omega}{4\pi^{2}}\frac{\omega_{0}^{2}}{2\pi}\nonumber \\
 & \times\sum_{n}\left(\left|C(\omega,\theta_{n,\text{Cl}})\boldsymbol{n}_{\bot}-\boldsymbol{D}_{\bot}(\omega,\theta_{n,\text{Cl}})\right|^{2}\right)d\omega d\varphi,\label{eq:classicalpower}
\end{flalign}
where we used the delta function to integrate over the angle $\theta$
such that 
\begin{equation}
\theta_{n,\text{Cl}}=\frac{1}{\gamma}\sqrt{\frac{2\gamma^{2}\omega_{0}n}{\omega}-\left(1+\gamma^{2}\left\langle \boldsymbol{v}_{\perp}^{2}\right\rangle \right)}.\label{eq:classicalangle}
\end{equation}
\begin{figure}[t]
\includegraphics[width=1\columnwidth]{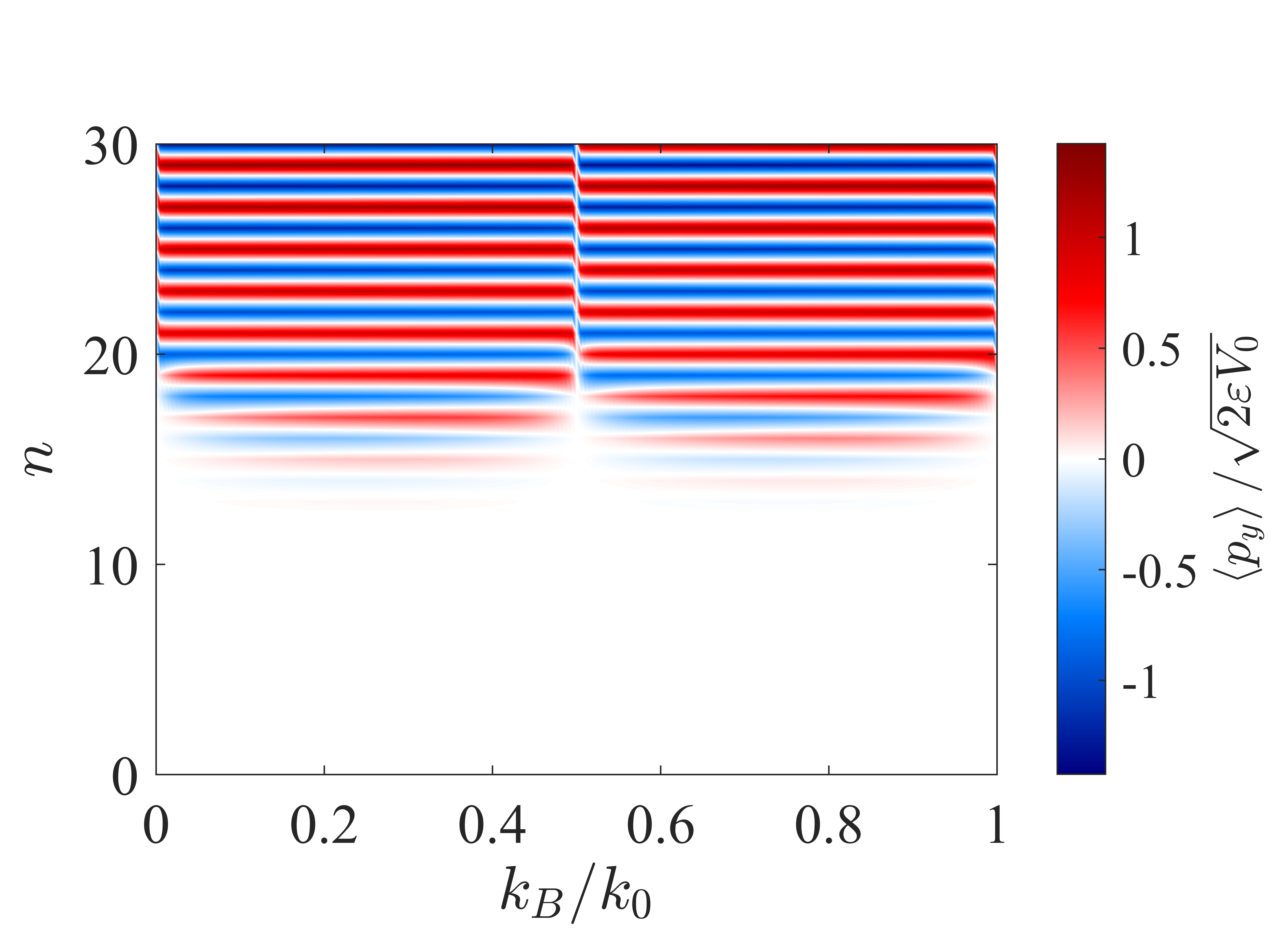} \caption{A plot of the expectation value of $p_{y}$ as function of the quantum
numbers $k_{B}$ and $n$ for a 100 MeV positron, corresponding to
figure 1.\label{fig:figbloch}}
\end{figure}
Note that for the sake of convenience we have introduced here a ``classical''
emission probability even though such a quantity has a meaning only
within the quantum theory. By performing the appropriate substitutions
of $\omega\rightarrow\omega'$ and by putting in the front factors
as in Eq. (\ref{eq:Baier formula}), we can obtain the semi-classical
version of the result in the form 
\begin{align}
\omega dW & =\frac{e^{2}\omega'}{4\pi^{2}}\frac{\omega_{0}^{2}}{2\pi}\left(\frac{\varepsilon^{2}+\varepsilon'^{2}}{2\varepsilon^{2}}\left|C(\omega',\theta_{n})\boldsymbol{n}_{\bot}-\boldsymbol{D}_{\bot}(\omega',\theta_{n})\right|^{2}\right)\nonumber \\
 & \left.+\frac{\omega^{2}m^{2}}{2\varepsilon^{4}}|C(\omega',\theta_{n})|^{2}\right)d\omega d\varphi,\label{eq:baierpower}
\end{align}
where 
\begin{equation}
\theta_{n}=\frac{1}{\gamma}\sqrt{\frac{2\gamma^{2}\omega_{0}n}{\omega'}-\left(1+\gamma^{2}\left\langle \boldsymbol{v}_{\perp}^{2}\right\rangle \right)}.\label{eq:angle}
\end{equation}
The threshold is therefore found to be at

\begin{equation}
\omega_{\text{th,B}}=\varepsilon\frac{2\gamma\omega_{0}n}{m\left(1+\gamma^{2}\left\langle \boldsymbol{v}_{\perp}^{2}\right\rangle \right)+2\gamma\omega_{0}n}.\label{eq:baierthreshold}
\end{equation}
\begin{figure}
\includegraphics[width=1\columnwidth]{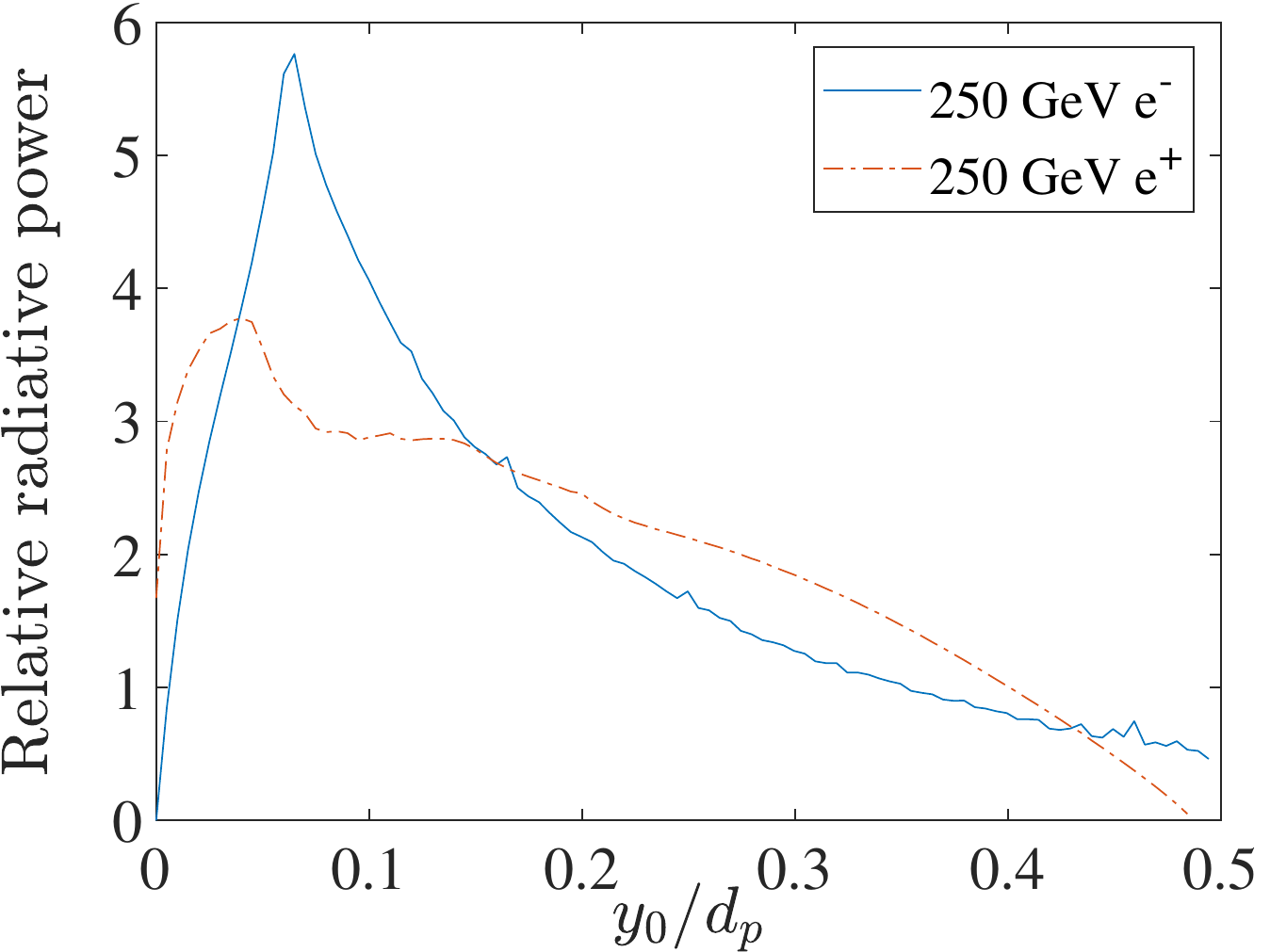}

\caption{The total power, relative to the average, of a 250 GeV electron/positron
incident at the position $y_{0}$ with zero angle.\label{fig:relpower}}

\end{figure}

\section{Discussion of results\label{sec:Discussion-of-results}}

In Fig. (\ref{fig:figpot}) we show the energy bands for a positron
with 100 MeV in the Doyle-Turner potential describing the (110) planes
of Silicon. It is noted that the energy $E_{n}$ of bound states with
small values of $n$ is almost independent of $k_{B}$. In Fig. (\ref{fig:figbloch})
we show a plot of the expectation value of the transverse momentum
$p_{y}$ and we see that the bound states, with $E_{n}<V_{0}$ have
$\left\langle p_{y}\right\rangle \simeq0$, and for the states above
the barrier, the quantity $\left\langle p_{y}\right\rangle $ steadily
increases with $E_{n}$. Also, it depends on $k_{B}$ in a way such
that half of the FBZ describes particles going to the left and the
other half particles going to the right.
\begin{figure}[t]
\includegraphics[width=1\columnwidth]{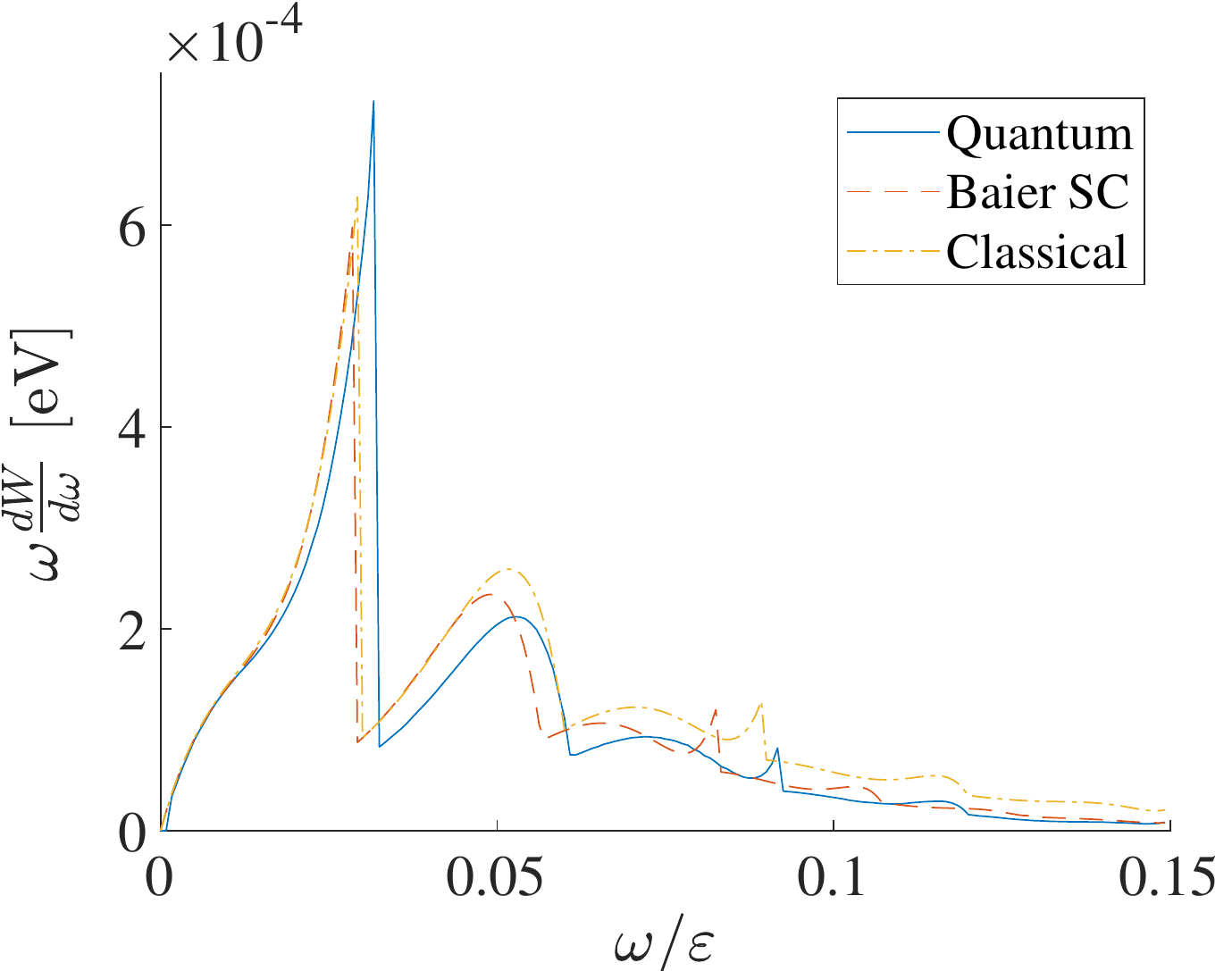} \caption{The differential power (probability per unit time multiplied by the
photon energy) for the case of a 20 GeV electron in the state $n=20$
between the (110) planes of Silicon. The label `Quantum' is the calculation
carried out by employing the wave functions found in this paper, `Baier
SC' is the semi-classical method of Baier et al. for a particle with
the same transverse mechanical energy and `Classical' is using the
classical formula for radiation emission.\label{fig:fig20GeVelec}}
\end{figure}
In Fig. (\ref{fig:relpower}) we show the total power of a 250 GeV electron/positron, calculated with
the LCFA, depending on the initial position $y_{0}$, when assuming
the beam angular divergence is negligible, i.e. much smaller than
the critical angle $\theta_{c}$. What can be seen from this figure
is that in both cases of the electron and positron, particles which
start out close to the plane (here $y_{0}=0$) have a larger radiation
power. However for positrons, as they are repelled from the planes,
this means a large oscillation amplitude, and therefore large $n$
quantum number, while for electrons, which are attracted to the planes,
starting close to the plane means a small oscillation amplitude and
therefore small $n$ quantum numbers radiate more for electrons. This,
combined with the fact that electrons will generally have an emission
spectrum distributed around larger photon energies, implies that all
the quantum effects mentioned in the introduction may become important.
In Fig. (\ref{fig:fig20GeVelec}) we show the calculation of the
photon emission spectra for a 20 GeV electron in the state with $n=20$.
This value of $n$ was picked as states around this value of $n$
have the largest rate of emission. 
In the particular case seen in Fig. (\ref{fig:fig20GeVelec}), we
can see the influence of the quantum effects in the following way. The fact that the classical calculation
differs from the semi-classical method of Baier et al. means that
the effects of spin and of photon recoil are present, i.e., that $\chi$
is large enough that these effects are sizable. At the same time,
we see that the calculations presented here differ from the semi-classical
of Baier et al. because the quantum number, $n=20$, of this state
is not large enough that the quantization of the motion can be completely
neglected. In addition, we have that $\xi$ is on the order of unity,
and therefore one can apply neither the constant field approximation
nor the dipole approximation. We see however in Fig. (\ref{fig:fig20GeVelec})
that the overall level of the semi-classical spectrum falls together
with the quantum results obtained here, while the most noticeable
difference is in the position of the thresholds, which should also
be clear from Eq. (\ref{eq:quantumthreshold}) and Eq. (\ref{eq:baierthreshold}),
see \citep{PhysRevA.98.022131,raicher2018semi} where this is also
found for different field configurations. Here, it is seen that the
threshold depends critically on the difference in energy between the
quantized levels in the transverse potential, information which is
not contained in the semi-classical method. In an experiment one would
however only obtain an average over the spectra corresponding to different
states, and therefore these details would likely be washed out. 
\begin{figure}[t]
\includegraphics[width=1\columnwidth]{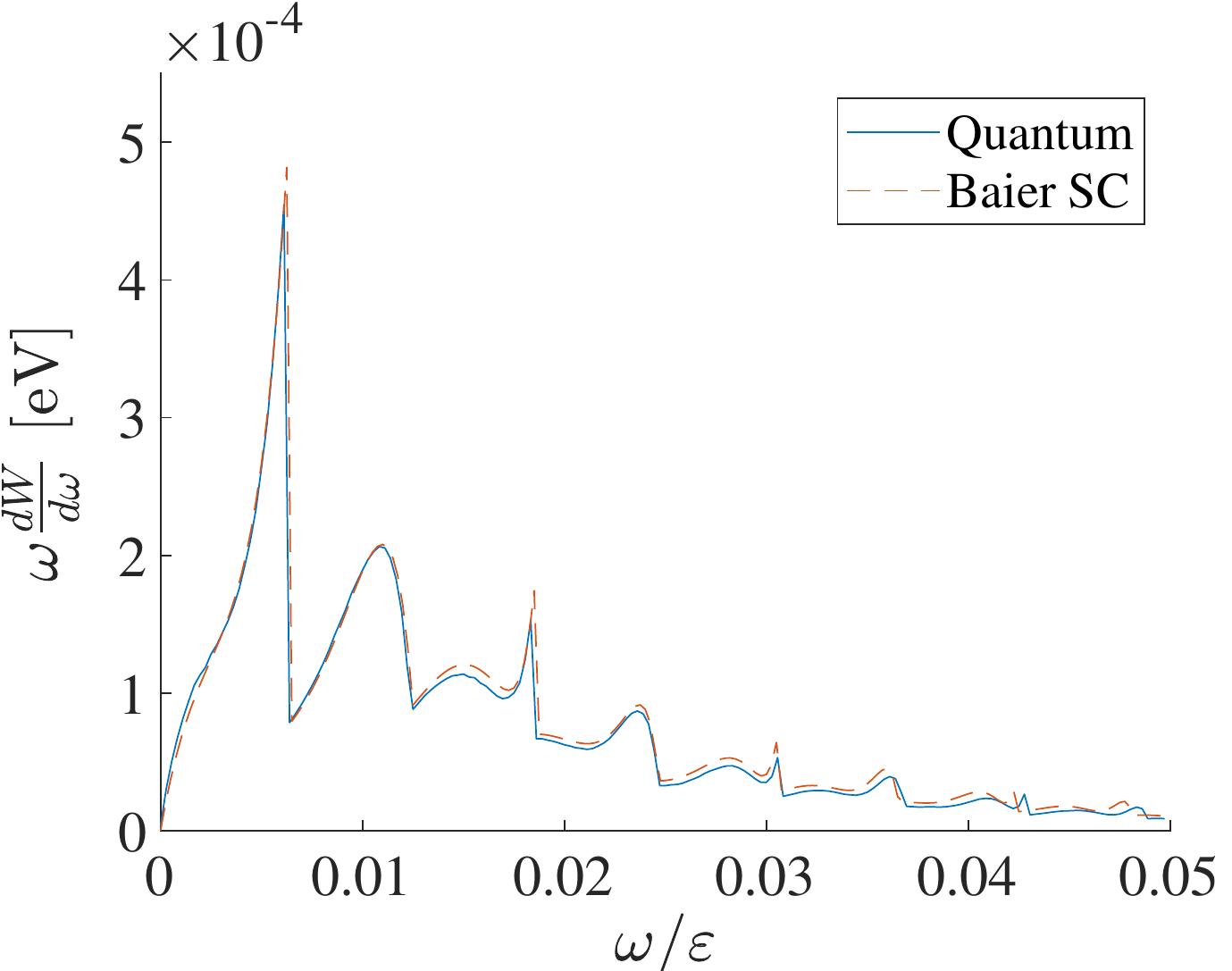} \caption{The differential power (probability per unit time multiplied by the
photon energy) for the case of a 10 GeV positron in the state $n=150$
between the (110) planes of Silicon. The label `Quantum' is the calculation
carried out by employing the wave functions found in this paper, `Baier
SC' is the semi-classical method of Baier et al. for a particle with
the same transverse mechanical energy.\label{fig:fig10GeVpos}}
\end{figure}
Definitive distinction could however be observed if the emitted photon
energy $\omega$ could be measured along with the emission angles
$\theta$ and $\varphi$. This can be seen from Eq. (\ref{eq:Quantumangle})
and Eq. (\ref{eq:angle}), which show that also the emission angle
depends on the level spacing. In addition, a qualitative difference
appears: the semi-classical treatment predicts that $\theta$ depends
only on $\omega$, while in the full calculation presented here, there
is also a dependence on $\varphi$ when $\omega\sim\varepsilon-\omega$
i.e. when $\chi$ is no longer small. In Fig. (\ref{fig:fig10GeVpos})
we show that while the same differences of the position of the thresholds
can be seen here, the effect is very small for positrons with larger
values of $n$ typically, as explained. However if we picked a positron
in a lower lying state, differences in the spectrum comparable to
those seen in Fig. (\ref{fig:fig20GeVelec}) would be seen, however
such low lying states have a low total radiated power for positrons,
and therefore do not contribute much to the spectrum when averaged over initial conditions.
\begin{figure}[t]
\includegraphics[width=1\columnwidth]{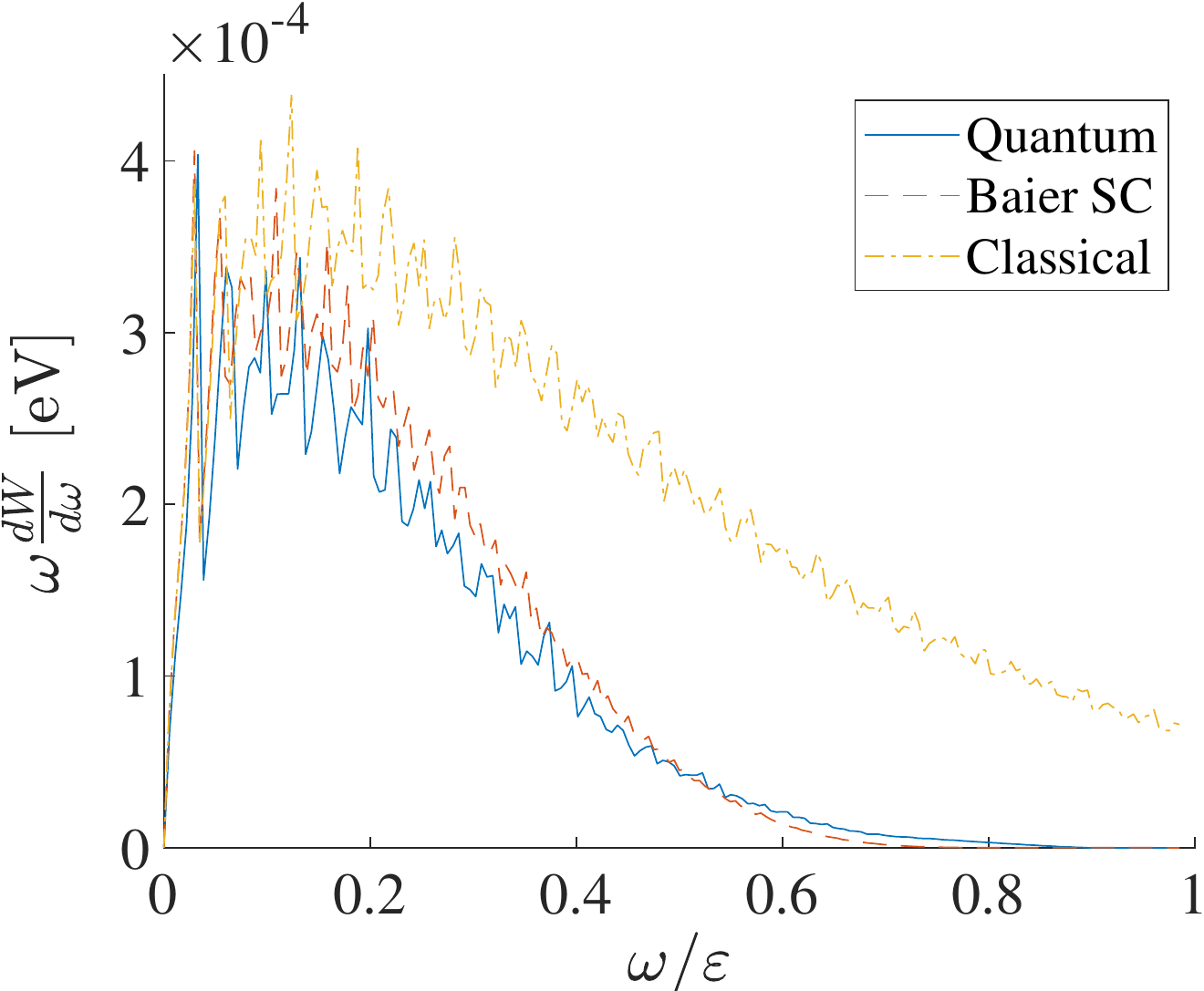} \caption{The differential power (probability per unit time multiplied by the
photon energy) for the case of a 250 GeV electron in the state $n=45$
between the (110) planes of Silicon. The label `Quantum' is the calculation
carried out by employing the wave functions found in this paper, `Baier
SC' is the semi-classical method of Baier et al. for a particle with
the same transverse mechanical energy, `Classical' is using the classical
formula for radiation emission, `LCFA' is the locally constant field
approximation and `BH' is the Bethe Heitler bremsstrahlung in amorphous
Silicon.\label{fig:fig250GeVelec}}
\end{figure}
 In Fig. (\ref{fig:fig250GeVelec}) we show the calculation for a
$250$ GeV electron in the $n=45$ state, which corresponds to the
particles with the largest power, i.e. the peak in Fig. (\ref{fig:relpower})
for electrons. We compare to the the semi-classical, classical and
the locally constant field approximation (LCFA) model. As expected,
the LCFA approximation agrees with the semi-classical for large photon
energies, where the formation length is short. It is seen that the
classical calculation fails for large photon energies, as expected.
For the very high energy part of the spectrum it is also observed
that the full quantum calculation is different from the semi-classical
result.  From Eq. (\ref{eq:quantumthreshold}) it can be seen that
larger values of the difference $E_{n_{i}}-E_{n_{f}}$ imply a larger
value of the emitted energy threshold. Therefore, the high-energy
part of the photon spectrum arises from transitions where the electron
goes from the initial state with $n=45$ to a state with a low value
of the quantum number $n$. Now, in the semi-classical method the
wave function of both initial and final states are approximated in
a way which is only valid when the corresponding discrete quantum
numbers are large. Since this is not the case for the final state,
the semi-classical result differs from the full quantum one. In Fig.
(\ref{fig:fig1TeVelec}) the same effect is seen for the case of a
$1$ TeV electron in the $n=120$ state initially. In this case the
effect is more pronounced, and far above the level of the Bethe-Heitler
bremsstrahlung which is approximately $\omega dW_{\text{BH}}/d\omega\simeq2.8\times10^{-6}$
eV. Therefore, if one carries out a precise measurement of the high-energy
part of the spectrum, the discrepancy between the two models could
be experimentally tested. In Fig. (\ref{fig:fig1TeVpos}) we show
the spectra for a 1 TeV positron with a large $n$ quantum number
and in this case, the semi-classical method works well.

\section{Conclusion\label{sec:Conclusion}}

\begin{figure}[t]
\includegraphics[width=1\columnwidth]{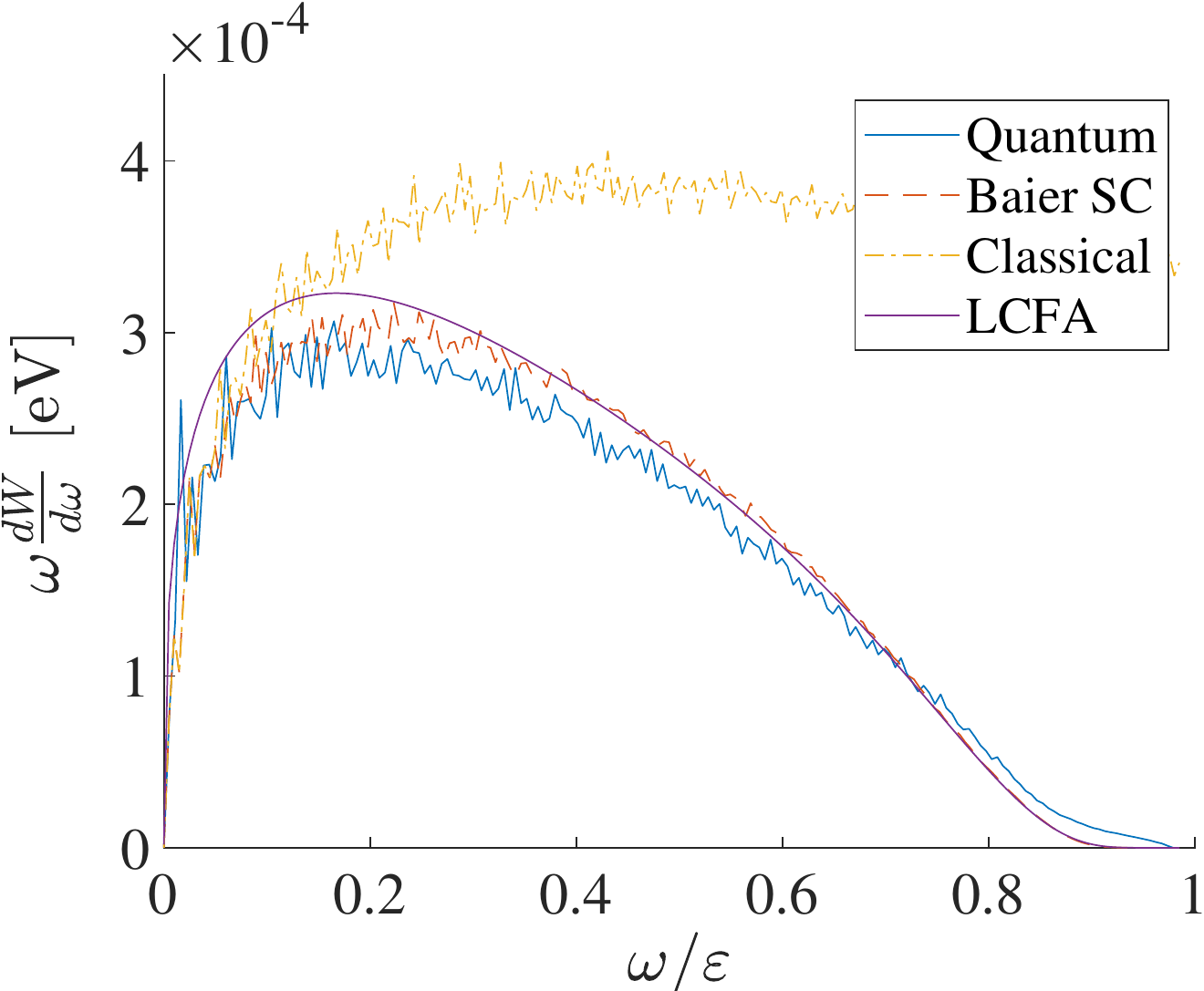} \caption{The differential power (probability per unit time multiplied by the
photon energy) for the case of a 1 TeV electron in the state $n=120$
between the (110) planes of Silicon. The labels have the same meaning
as in Fig. (\ref{fig:fig250GeVelec}).\label{fig:fig1TeVelec}}
\end{figure}
In conclusion, we have shown how to find approximate solutions of
the Dirac equation for describing the motion of relativistic electrons
and positrons in a periodic potential which depends on one transverse
coordinate, as compared to the direction of the largest momentum (here
indicated as the $x$-direction). We have shown how to calculate the
emission rate of a single photon from transitions between the corresponding
quantum states exactly, that is, without the use
of neither the dipole approximation nor the locally constant field
approximation. Therefore, we have been able to calculate
single-photon planar channeling radiation, with all relevant quantum
effects included: the effects of electron/positron spin and photon
recoil during the emission, which the semi-classical method also incorporates,
but also the effects of the quantization of the transverse motion.
For planar channeling, and in particular for positrons, we saw that
the semi-classical approximation of Baier et al. (beyond the locally
constant field approximation) is accurate in describing the energy
distribution of the emitted photon, when integrated over angles. For
electrons, differences are more noticeable. For low electron energies,
a clear experimental measurement of such differences would require
angular resolution of the emitted photons along with their energy.
However, for higher-energy electrons a difference could potentially
be detected experimentally, even in the angularly integrated emission
spectrum for emitted photons with high energy.

\section{Acknowledgments}

T. W. is supported by the Alexander von Humboldt-Stiftung apart from
the initial part of the project where funding was provided by the
VILLUM FONDEN (research grant VKR023371).

\begin{figure}[H]
\includegraphics[width=1\columnwidth]{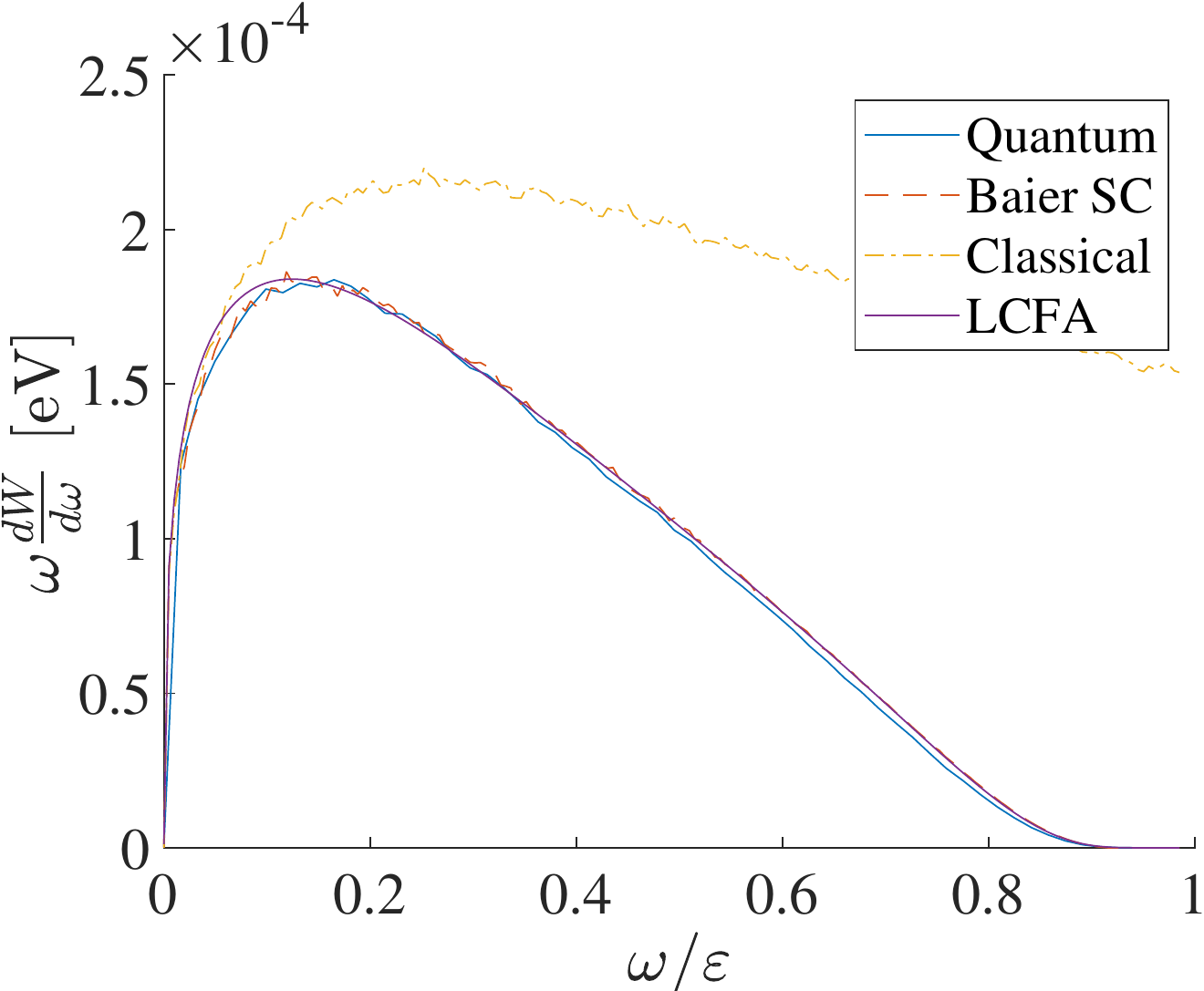}

\caption{The differential power (probability per unit time multiplied by the
photon energy) for the case of a 1 TeV positron in the state $n=1512$
between the (110) planes of Silicon. The labels have the same meaning
as in Fig. (\ref{fig:fig250GeVelec}).\label{fig:fig1TeVpos}}
\end{figure}
\newpage{}

\begin{widetext}

\section*{Appendix A}

The general (unnormalized) solution to the Dirac equation with potential
energy $V(\boldsymbol{r})=-e\varphi(\boldsymbol{r})$ can be written
as

\begin{equation}
\psi(\boldsymbol{r},t)=e^{-i\varepsilon t}\left(\begin{array}{c}
\phi(\boldsymbol{r})\\
\chi(\boldsymbol{r})
\end{array}\right)
\end{equation}
The Dirac equation then becomes

\begin{equation}
\left(\varepsilon+e\varphi-m\right)\phi(\boldsymbol{r})=\boldsymbol{\sigma}\cdot\hat{\boldsymbol{p}}\chi(\boldsymbol{r}),\label{eq:Dirac1}
\end{equation}

\begin{equation}
\left(\varepsilon+e\varphi+m\right)\chi(\boldsymbol{r})=\boldsymbol{\sigma}\cdot\hat{\boldsymbol{p}}\phi(\boldsymbol{r}).\label{eq:Dirac2}
\end{equation}
If we insert $\chi(\boldsymbol{r})$ from Eq. (\ref{eq:Dirac2}) in
Eq. (\ref{eq:Dirac1}) we can obtain an equation for $\phi(\boldsymbol{r})$.

\begin{flalign}
\left(\varepsilon+e\varphi-m\right)\phi(\boldsymbol{r}) & =\boldsymbol{\sigma}\cdot\hat{\boldsymbol{p}}\frac{1}{\left(\varepsilon+e\varphi+m\right)}\boldsymbol{\sigma}\cdot\hat{\boldsymbol{p}}\phi(\boldsymbol{r})\nonumber \\
 & =\frac{1}{\left(\varepsilon+e\varphi+m\right)}\left[\boldsymbol{\sigma}\cdot\hat{\boldsymbol{p}}\right]^{2}\phi(\boldsymbol{r})-i\frac{e\boldsymbol{\sigma}\cdot\boldsymbol{E}}{\left(\varepsilon+e\varphi+m\right)^{2}}\boldsymbol{\sigma}\cdot\hat{\boldsymbol{p}}\phi(\boldsymbol{r})\nonumber \\
 & =\frac{1}{\left(\varepsilon+e\varphi+m\right)}\hat{\boldsymbol{p}}^{2}\phi(\boldsymbol{r})-i\frac{e\boldsymbol{\sigma}\cdot\boldsymbol{E}}{\left(\varepsilon+e\varphi+m\right)^{2}}\boldsymbol{\sigma}\cdot\hat{\boldsymbol{p}}\phi(\boldsymbol{r})
\end{flalign}
Now let us multiply with $\left(\varepsilon+e\varphi+m\right)$, then

\begin{flalign}
\hat{\boldsymbol{p}}^{2}\phi(\boldsymbol{r})-i\frac{e\boldsymbol{\sigma}\cdot\boldsymbol{E}}{\varepsilon+e\varphi+m}\boldsymbol{\sigma}\cdot\hat{\boldsymbol{p}}\phi(\boldsymbol{r}) & =\left(\varepsilon+e\varphi+m\right)\left(\varepsilon+e\varphi-m\right)\phi(\boldsymbol{r})\nonumber \\
 & =\left[\left(\varepsilon+e\varphi\right)^{2}-m^{2}\right]\phi(\boldsymbol{r})\nonumber \\
 & =\left[\varepsilon^{2}+2\varepsilon e\varphi+e^{2}\varphi^{2}-m^{2}\right]\phi(\boldsymbol{r})
\end{flalign}
now if we neglect the spin-field interaction terms and the field squared
term $e^{2}\varphi^{2}$ this becomes

\begin{equation}
\left[\frac{1}{2\varepsilon}\hat{\boldsymbol{p}}^{2}-e\varphi\right]\phi(\boldsymbol{r})=\frac{\varepsilon^{2}-m^{2}}{2\varepsilon}\phi(\boldsymbol{r})
\end{equation}
The electron solution is then

\begin{equation}
\psi(\boldsymbol{r},t)=e^{-i\varepsilon t}\left(\begin{array}{c}
\phi(\boldsymbol{r})\\
\frac{\boldsymbol{\sigma}\cdot\hat{\boldsymbol{p}}}{\varepsilon-V(\boldsymbol{r})+m}\phi(\boldsymbol{r})
\end{array}\right).
\end{equation}
The expression from Eq. \eqref{eq:Uspinor} is then obtained by setting
$\phi(\boldsymbol{r})=\boldsymbol{s}e^{i(p_{x}x+p_{z}z)}I(y)$ and
expanding $\left(\varepsilon-V(\boldsymbol{r})+m\right)^{-1}\simeq\left(\varepsilon+m\right)^{-1}\left(1+\frac{V(\boldsymbol{r})}{\varepsilon+m}\right)$,
and keeping this correction term with the potential only on the $\sigma_{x}p_{x}$
term, as here this correction yields the leading order in $\xi/\gamma$
on $\sigma_{x}$.

\section*{Appendix B}

The electron state can be written as (putting back in the volume factor)

\begin{equation}
\psi_{p,s}(x)=\frac{1}{\sqrt{2\varepsilon V}}e^{-i\varepsilon_{n}t}e^{i(p_{x}x+k_{B}y+p_{z}z)}\sum_{j}c_{j}\boldsymbol{S}_{j}e^{ijk_{0}y},\label{eq:electronwavefunc-1}
\end{equation}
where

\begin{equation}
\boldsymbol{S}_{j}=\sqrt{\varepsilon+m}\left(\begin{array}{c}
\boldsymbol{s}\\
\frac{\boldsymbol{\sigma}\cdot\boldsymbol{p}_{j}}{\varepsilon+m}\boldsymbol{s}
\end{array}\right),\label{eq:bigS-1}
\end{equation}
where $\boldsymbol{p}_{j}=(p_{x}+E_{n}-\frac{(jk_{0}+k_{B})^{2}}{2\varepsilon},jk_{0}+k_{B},p_{z})$
and then

\begin{equation}
\int\psi_{p'}^{\dagger}\psi_{p}dV=\frac{1}{2V\sqrt{\varepsilon'\varepsilon}}(2\pi)^{3}\delta(p_{x}-p_{x}')\delta(p_{z}-p_{z}')\sum_{j,j'}c_{j}(p)c_{j'}^{*}(p')\boldsymbol{S}_{j'}^{'\dagger}\boldsymbol{S}_{j}\delta(k_{B}-k_{B}'+(j-j')k_{0}).
\end{equation}
Explicitly we have that $c_{j}=c_{j}(\varepsilon,k_{B},n)$. Now since
both $k_{B}$ and $k_{B}'$ obey that $0\leq k_{B}<k_{0}$ we have
that $-k_{0}<k_{B}-k_{B}'<k_{0}$ and therefore $k_{B}-k_{B}'$ can
never be an integer value of $k_{0}$ unless $k_{B}-k_{B}'=0$, and
therefore we can write

\begin{equation}
\delta(k_{B}-k_{B}'+(j-j')k_{0})=\delta(k_{B}-k_{B}')\delta_{j,j'}
\end{equation}

\begin{equation}
\int\psi_{p'}^{\dagger}\psi_{p}dV=\frac{1}{2V\sqrt{\varepsilon'\varepsilon}}(2\pi)^{3}\delta(p_{x}-p_{x}')\delta(p_{z}-p_{z}')\delta(k_{B}-k_{B}')\sum_{j}c_{j}(\varepsilon,k_{B},n)c_{j}^{*}(\varepsilon,k_{B},n')\boldsymbol{S}_{j}^{'\dagger}\boldsymbol{S}_{j}
\end{equation}
However the vector $\boldsymbol{c}$ is a normalized ($|\boldsymbol{c}|=1$),
eigenvector of a hermitian matrix and the vectors corresponding to
$n$ and $n'$ have different eigenvalues of this matrix, and are
therefore orthogonal, so

\begin{equation}
\int\psi_{p'}^{\dagger}\psi_{p}dV=\frac{1}{2V\sqrt{\varepsilon'\varepsilon}}(2\pi)^{3}\delta(p_{x}-p_{x}')\delta(p_{z}-p_{z}')\delta(k_{B}-k_{B}')\delta_{n,n'}\sum_{j}|c_{j}|^{2}\boldsymbol{S}_{j}^{'\dagger}\boldsymbol{S}_{j}.
\end{equation}
Now consider

\begin{equation}
\boldsymbol{S}_{j}^{'\dagger}\boldsymbol{S}_{j}=(\varepsilon+m)\left(\boldsymbol{s}^{'\dagger}\boldsymbol{s}+\boldsymbol{s}^{'\dagger}\frac{\boldsymbol{\sigma}\cdot\boldsymbol{p}_{j}}{\varepsilon+m}\frac{\boldsymbol{\sigma}\cdot\boldsymbol{p}_{j}}{\varepsilon+m}\boldsymbol{s}\right)=\boldsymbol{s}^{'\dagger}\boldsymbol{s}\left[(\varepsilon+m)+\frac{\boldsymbol{p}_{j}^{2}}{\varepsilon+m}\right],
\end{equation}
and therefore 
\begin{equation}
\sum_{j}|c_{j}|^{2}\boldsymbol{S}_{j}^{'\dagger}\boldsymbol{S}_{j}\simeq2\varepsilon\delta_{s',s}.
\end{equation}
There $\simeq$ refers only to the normalization. The states are exactly
orthogonal, but in the normalization we neglect corrections which
are suppressed by at least $\xi/\gamma$ compared to leading order.
So finally

\begin{equation}
\int\psi_{p'}^{\dagger}\psi_{p}dV=\frac{(2\pi)^{3}}{V}\delta(p_{x}-p_{x}')\delta(p_{z}-p_{z}')\delta(k_{B}-k_{B}')\delta_{n,n'}\delta_{s',s}.
\end{equation}

\section*{Appendix C}

Inserting our wave-functions into the expression from the paper, only
the $y$ component is non-trivial:

\begin{alignat}{1}
\left\langle p_{y}\right\rangle  & =\frac{1}{2\varepsilon L}\int\left(e^{-ik_{B}y}\sum_{l}c_{l}^{*}\boldsymbol{S}_{l}^{\dagger}e^{-ilk_{0}y}\right)\left(-i\frac{d}{dy}\right)\left(e^{ik_{B}y}\sum_{j}c_{j}\boldsymbol{S}_{j}e^{ijk_{0}y}\right)dy\nonumber \\
 & =\frac{1}{2\varepsilon L}\int\left(\sum_{l}c_{l}^{*}\boldsymbol{S}_{l}^{\dagger}e^{-i(lk_{0}+k_{B})y}\right)\left(-i\frac{d}{dy}\right)\left(\sum_{j}c_{j}\boldsymbol{S}_{j}e^{i(jk_{0}+k_{B})y}\right)dy\nonumber \\
 & =\frac{1}{2\varepsilon L}\int\left(\sum_{l}c_{l}^{*}\boldsymbol{S}_{l}^{\dagger}e^{-i(lk_{0}+k_{B})y}\right)\left(\sum_{j}(jk_{0}+k_{B})c_{j}\boldsymbol{S}_{j}e^{i(jk_{0}+k_{B})y}\right)dy\nonumber \\
 & =\frac{1}{2\varepsilon L}\int\left(\sum_{j,l}(jk_{0}+k_{B})c_{j}c_{l}^{*}\boldsymbol{S}_{l}^{\dagger}\boldsymbol{S}_{j}e^{i(j-l)k_{0}y}\right)dy
\end{alignat}
Now due to periodicity of the integrand we have that

\begin{equation}
\frac{1}{L}\int_{-L/2}^{L/2}e^{i(j-l)k_{0}y}dy=\frac{1}{d_{p}}\int_{-d_{p}/2}^{d_{p}/2}e^{i(j-l)k_{0}y}dy=\text{sinc}\left((j-l)\pi\right)=\delta_{j,l}
\end{equation}
and so

\begin{align}
\left\langle p_{y}\right\rangle  & =\frac{1}{2\varepsilon}\left(\sum_{j}\left|c_{j}\right|^{2}(jk_{0}+k_{B})\boldsymbol{S}_{j}^{\dagger}\boldsymbol{S}_{j}\right)\nonumber \\
 & =\frac{1}{2\varepsilon}\left(\sum_{j}\left|c_{j}\right|^{2}(jk_{0}+k_{B})\left((\varepsilon+m)+\frac{\boldsymbol{p}_{j}^{2}}{\varepsilon+m}\right)\right).
\end{align}
and setting $p_{z}=0$,

\begin{flalign}
\boldsymbol{p}_{j}^{2} & =\left(p_{x}+E_{n}-\frac{(jk_{0}+k_{B})^{2}}{2\varepsilon}\right)^{2}+(jk_{0}+k_{B})^{2}\simeq p_{x}^{2},
\end{flalign}
and therefore

\begin{align}
\frac{1}{2\varepsilon}\left((\varepsilon+m)+\frac{\boldsymbol{p}_{j}^{2}}{\varepsilon+m}\right) & \simeq1.
\end{align}
And so within our level of approximation we have that

\begin{equation}
\left\langle p_{y}\right\rangle \simeq\sum_{j}\left|c_{j}\right|^{2}(jk_{0}+k_{B}).
\end{equation}

\section*{Appendix D}

Starting from

\begin{align}
S_{fi}^{(1)} & =ie\sqrt{\frac{4\pi}{2\omega}}\frac{1}{2\sqrt{\varepsilon_{f}\varepsilon_{i}}}(2\pi)^{3}\delta(p_{x,i}-k_{x}-p_{x,f})\delta(p_{z,i}-k_{z}-p_{z,f})\nonumber \\
 & \times\delta(\varepsilon_{f}+\omega-\varepsilon_{i})\int_{-\infty}^{\infty}\bar{U}_{f}(y)\slashed{\epsilon}^{*}e^{i(k_{B,i}-k_{y}-K_{B,f})y}U_{i}(y)dy,
\end{align}
we insert the wave functions in terms of the plane wave expansion
to obtain

\begin{align}
S_{fi}^{(1)} & =ie\sqrt{\frac{4\pi}{2\omega}}\frac{1}{2\sqrt{\varepsilon_{f}\varepsilon_{i}}}(2\pi)^{3}\delta(p_{x,i}-k_{x}-p_{x,f})\delta(p_{z,i}-k_{z}-p_{z,f})\nonumber \\
 & \times\delta(\varepsilon_{f}+\omega-\varepsilon_{i})\nonumber \\
 & \times\sum_{j,l}\int_{-\infty}^{\infty}c_{l,f}^{*}c_{j,i}\boldsymbol{\bar{S}}_{l,f}\slashed{\epsilon}^{*}\boldsymbol{S}_{j,i}e^{i(k_{B,i}-k_{y}-k_{B,f})y}e^{i(j-l)k_{0}y}dy.
\end{align}
The quantity $F=\int_{-\infty}^{\infty}e^{i(k_{B,i}-k_{y}-k_{B,f})y}e^{i(j-l)k_{0}y}dy$
can be rewritten by exploiting that $e^{i(j-l)k_{0}y}$ is periodic,
and so we see that

\begin{align}
F & =\sum_{n=-\infty}^{\infty}\int_{-\frac{d_{p}}{2}+nd_{p}}^{\frac{d_{p}}{2}+nd_{p}}e^{i(k_{B,i}-k_{y}-k_{B,f})y}e^{i(j-l)k_{0}y}dy\label{eq:F1}
\end{align}
and change variable such that $y'=y-nd_{p}$,

\begin{align}
F & =\sum_{n=-\infty}^{\infty}\int_{-\frac{d_{p}}{2}}^{\frac{d_{p}}{2}}e^{i(k_{B,i}-k_{y}-k_{B,f})(y'+nd_{p})}e^{i(j-l)k_{0}y'}dy'\nonumber \\
 & =\sum_{n=-\infty}^{\infty}e^{ind_{p}\left(k_{B,i}-k_{y}-k_{B,f}\right)}\nonumber \\
 & \times\int_{-\frac{d_{p}}{2}}^{\frac{d_{p}}{2}}e^{i(k_{B,i}-k_{y}-k_{B,f}+(j-l)k_{0})y'}dy'\label{eq:F2}
\end{align}
Now the sum $\sum_{n=-\infty}^{\infty}e^{ind_{p}\left(k_{B,i}-k_{y}-k_{B,f}\right)}$
can be recognized as the Dirichlet kernel which can be replaced with
the Dirac comb $\sum_{n=-\infty}^{\infty}2\pi\delta(\left[k_{B,i}-k_{y}-k_{B,f}\right]d_{p}-2\pi n)$.
Only the delta function which has $k_{B,f}$ in the first Brillouin
zone will contribute, due to the fact that integration limit on $k_{B,f}$
is from $0$ to $k_{0}$, and therefore we must set $n=n_{B}$ such
that $0\leq k_{B,f}<k_{0}$. Therefore we may use that

\begin{align}
F & =2\pi\delta(\left[k_{B,i}-k_{y}-k_{B,f}\right]-n_{B}k_{0})\nonumber \\
 & \times\text{sinc}(\left[\pi n_{B}+(j-l)\pi\right]),\label{eq:Fsinc}
\end{align}
This simplifies the summation over $l$ as the sinc function means
that only the term obeying $n_{B}+(j-l)=0$ contributes. Therefore
the S-matrix element becomes

\begin{align}
S_{fi}^{(1)} & =ie\sqrt{\frac{4\pi}{2\omega}}\frac{1}{2\sqrt{\varepsilon_{f}\varepsilon_{i}}}(2\pi)^{4}\delta(p_{x,i}-k_{x}-p_{x,f})\delta(p_{z,i}-k_{z}-p_{z,f})\nonumber \\
 & \times\delta(\varepsilon_{f}+\omega-\varepsilon_{i})\delta(\left[k_{B,i}-k_{y}-k_{B,f}\right]-n_{B}k_{0})\nonumber \\
 & \times\sum_{j}c_{n_{B}+j,f}^{*}c_{j,i}\bar{\boldsymbol{S}}_{n_{B}+j,f}\slashed{\epsilon}^{*}\boldsymbol{S}_{j,i}.
\end{align}
Now a useful expression for the quantity $\bar{\boldsymbol{S}}_{n_{B}+j,f}\slashed{\epsilon}^{*}\boldsymbol{S}_{j,i}$
may be derived. We will momentarily suppress the $j$ index, and so

\begin{align}
-\bar{\boldsymbol{S}}_{f}\slashed{\epsilon}^{*}\boldsymbol{S}_{i} & =\sqrt{(\varepsilon_{f}+m)(\varepsilon_{i}+m)}\left(\begin{array}{cc}
\boldsymbol{s}_{f}^{\dagger} & \boldsymbol{s}_{f}^{\dagger}\frac{\boldsymbol{\sigma}\cdot\boldsymbol{p}_{f}}{\varepsilon_{f}+m}\end{array}\right)\left(\boldsymbol{\alpha}\cdot\boldsymbol{\epsilon}^{*}\right)\left(\begin{array}{c}
\boldsymbol{s}_{i}\\
\frac{\boldsymbol{\sigma}\cdot\boldsymbol{p}_{i}}{\varepsilon_{i}+m}\boldsymbol{s}_{i}
\end{array}\right)\nonumber \\
 & =\sqrt{(\varepsilon_{f}+m)(\varepsilon_{i}+m)}\left(\begin{array}{cc}
\boldsymbol{s}_{f}^{\dagger} & \boldsymbol{s}_{f}^{\dagger}\frac{\boldsymbol{\sigma}\cdot\boldsymbol{p}_{f}}{\varepsilon_{f}+m}\end{array}\right)\left(\begin{array}{cc}
0 & \boldsymbol{\sigma}\cdot\boldsymbol{\epsilon}^{*}\\
\boldsymbol{\sigma}\cdot\boldsymbol{\epsilon}^{*} & 0
\end{array}\right)\left(\begin{array}{c}
\boldsymbol{s}_{i}\\
\frac{\boldsymbol{\sigma}\cdot\boldsymbol{p}_{i}}{\varepsilon_{i}+m}\boldsymbol{s}_{i}
\end{array}\right)\nonumber \\
 & =\sqrt{(\varepsilon_{f}+m)(\varepsilon_{i}+m)}\left(\begin{array}{cc}
\boldsymbol{s}_{f}^{\dagger} & \boldsymbol{s}_{f}^{\dagger}\frac{\boldsymbol{\sigma}\cdot\boldsymbol{p}_{f}}{\varepsilon_{f}+m}\end{array}\right)\left(\begin{array}{c}
\boldsymbol{\sigma}\cdot\boldsymbol{\epsilon}^{*}\frac{\boldsymbol{\sigma}\cdot\boldsymbol{p}_{i}}{\varepsilon_{i}+m}\boldsymbol{s}_{i}\\
\boldsymbol{\sigma}\cdot\boldsymbol{\epsilon}^{*}\boldsymbol{s}_{i}
\end{array}\right)\nonumber \\
 & =\sqrt{(\varepsilon_{f}+m)(\varepsilon_{i}+m)}\boldsymbol{s}_{f}^{\dagger}\left(\boldsymbol{\sigma}\cdot\boldsymbol{\epsilon}^{*}\frac{\boldsymbol{\sigma}\cdot\boldsymbol{p}_{i}}{\varepsilon_{i}+m}+\frac{\boldsymbol{\sigma}\cdot\boldsymbol{p}_{f}}{\varepsilon_{f}+m}\boldsymbol{\sigma}\cdot\boldsymbol{\epsilon}^{*}\right)\boldsymbol{s}_{i}\nonumber \\
 & =\sqrt{(\varepsilon_{f}+m)(\varepsilon_{i}+m)}\boldsymbol{s}_{f}^{\dagger}\left(\frac{1}{\varepsilon_{i}+m}\left[\boldsymbol{\epsilon}^{*}\cdot\boldsymbol{p}_{i}+i\boldsymbol{\sigma}\cdot\left(\boldsymbol{\epsilon}^{*}\times\boldsymbol{p}_{i}\right)\right]+\frac{1}{\varepsilon_{f}+m}\left[\boldsymbol{p}_{f}\cdot\boldsymbol{\epsilon}^{*}+i\boldsymbol{\sigma}\cdot\left(\boldsymbol{p}_{f}\times\boldsymbol{\epsilon}^{*}\right)\right]\right)\boldsymbol{s}_{i}\\
 & =\sqrt{(\varepsilon_{f}+m)(\varepsilon_{i}+m)}\boldsymbol{s}_{f}^{\dagger}\left[\boldsymbol{\epsilon}^{*}\cdot\boldsymbol{A}+i\boldsymbol{\sigma}\cdot\boldsymbol{B}\right]\boldsymbol{s}_{i}\nonumber 
\end{align}
where we have defined 
\begin{equation}
\boldsymbol{A}=\frac{1}{\varepsilon_{i}+m}\boldsymbol{p}_{i}+\frac{1}{\varepsilon_{f}+m}\boldsymbol{p}_{f},
\end{equation}

\begin{equation}
\boldsymbol{B}=\boldsymbol{\epsilon}^{*}\times\left(\frac{\boldsymbol{p}_{i}}{\varepsilon_{i}+m}-\frac{\boldsymbol{p}_{f}}{\varepsilon_{f}+m}\right).
\end{equation}

\end{widetext}

\bibliography{biblio}

\end{document}